\documentclass[aps,prl,twocolumn,showpacs,preprintnumbers,amsmath,amssymb,revsymb]{revtex4}
\usepackage{graphicx}
\usepackage{color} 
\usepackage{dcolumn}
\usepackage{bm}

\def\braket#1{\mathinner{\langle{#1}\rangle}}
\def\bra#1{\mathinner{\langle{#1}|}}
\def\ket#1{\mathinner{|{#1}\rangle}}

\begin{document}

\title{Universal detector efficiency of a mesoscopic capacitor}

\preprint{}

\author{Simon~E.~Nigg}\email[Corresponding author: ]{simon.nigg@unige.ch}
\author{Markus~B\"uttiker}
\affiliation{D\'epartement de Physique Th\'eorique, Universit\'e de
  Gen\`eve, CH-1211 Gen\`eve 4, Switzerland}
\date{\today}

\begin{abstract}
We investigate theoretically a novel type of high frequency quantum detector
based on the mesoscopic capacitor recently realized by Gabelli
et al., [Science {\bf 313}, 499 (2006)], which consists of a quantum
dot connected via a single channel quantum point contact to a single lead. We show that
the state of a double quantum  dot charge qubit
capacitively coupled to this detector can be read out in the GHz
frequency regime with near
quantum limited efficiency. To leading order, the quantum efficiency
is found to be universal owing to the universality of the charge relaxation
resistance of the mesoscopic capacitor. 
\end{abstract}

\maketitle
The measurement problem is probably one of the oldest topics
in quantum physics, which is still of prime interest to researchers
nowadays. With the advent of mesoscopic physics, fundamental issues
related to Von Neumann's notion of the instantaneous wave
function collapse~\cite{VonNeumann:32} can now be addressed experimentally. Indeed it has recently become possible to engineer
systems in which parts of the measurement device are themselves
unambiguously quantum. In the weak coupling regime the dynamics of the
wave function collapse itself can be probed and sometimes even reversed~\cite{Katz:08-short,Korotkov:06}. Questions such as ``how long does it take to acquire the desired information ?'' and ``how fast does the
measurement decoher the state of the measured system ?'' become of
relevance. This is in particular true in the emergent field of quantum information
processing, where one wishes to both manipulate and read-out quantum
bits (qubits) with the highest possible efficiencies.

An important figure of merit of any quantum
detector is its Heisenberg efficiency. Loosely speaking it is
the ratio of how fast
to how invasive a given detector is. By ``fast'' we mean how
quickly two different states
of the measured system can be distinguished from one another and by
``invasive'' we mean how strong is the back-action of the detector onto the state of the measured system. The Heisenberg
uncertainty relation implies that one cannot acquire information about
the system faster than one dephases it during the measurement
process~\cite{Korotkov:01,Makhlin:01,Clerk:03,Pilgram:02}. Hence the
Heisenberg efficiency is bounded
from above. An important task is thus to find and characterize detectors which reach the
maximum allowed Heisenberg efficiency.

Several such systems have been described in the
literature. In the DC regime Refs.~\cite{Gurvitz:97,Aleiner:97,Clerk:04,Averin:05}
investigate the quantum point contact (QPC) detector. Refs.~\cite{Pilgram:02,Clerk:03} discuss
two terminal scattering detectors capacitively coupled to a double dot
charge qubit. In both cases, the average current through
the detector functions as a meter, since the electron transmission probability
is sensitive to the position of the charge in the qubit. Due to
$1/f$ noise DC
detectors are generically plagued by a large dephasing rate. To circumvent this, Schoelkopf et al.~\cite{Schoelkopf:98} introduced the radio-frequency
single-electron transistor (rf-SET). The idea there, is to measure the {\em
  damping} of an oscillator circuit in which the SET is
embedded. 

In this letter we present a novel quantum detector based on the
mesoscopic capacitor~\cite{Buttiker:93b}, which consists of a quantum
dot connected via a single channel QPC to a single lead. At
temperatures low compared with the charging energy, such a system
exhibits a universal~\cite{Buttiker:93b,Gabelli:06,Nigg:06} charge relaxation
resistance $R_q=h/(2e^2)$. We show that this system embedded in an LC tank circuit with impedance
$L$ and capacitance $C$, can be operated as a high frequency detector near
the quantum limit despite the presence of intrinsic dissipation. At the
resonance frequency $\omega_0=1/\sqrt{LC}$ we find to leading
order, a {\em universal} Heisenberg efficiency
\begin{equation}\label{eq:10}
\eta = \frac{L/C}{L/C+R_qZ_0}\,,
\end{equation}
where $Z_0$ is the characteristic impedance of the
transmission line connected to the tank circuit.
\begin{figure}[b]
\begin{center}
\includegraphics[width=.45\textwidth]{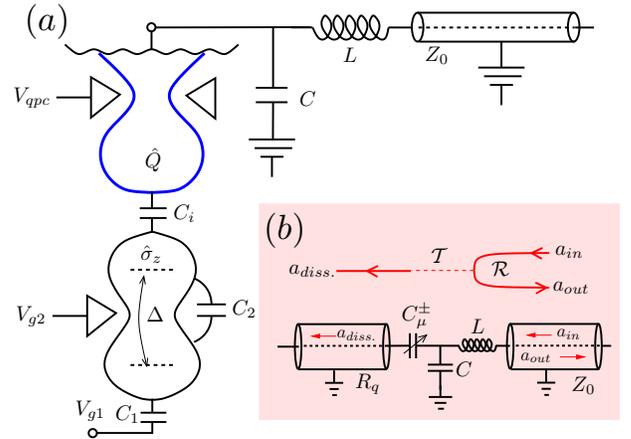}\caption{(Color online)
  Detector and qubit system (a). Equivalent circuit in the
  adiabatic approximation (b). Incoming photons are reflected
  ($\mathcal{R}$) and detected or dissipated ($\mathcal{T}$).
  \label{fig:1}}
\end{center}
\end{figure}

The system we consider is depicted in Fig.~\ref{fig:1} (a). Let us first
consider the system without the LC resonator and transmission
line. The part
being measured; the double dot charge qubit, plus the capacitive coupling term
are described by the Hamiltonian
\begin{equation}\label{eq:27}
H_{qb}=\frac{1}{2}\left(\epsilon\sigma_z+\Delta\sigma_x +\kappa\sigma_z\hat N\right)\quad\text{with}\quad\kappa=\frac{e^2}{C_i}\,.
\end{equation}
Here $\sigma_z=\ket{\uparrow}\bra{\uparrow}-\ket{\downarrow}\bra{\downarrow}$ and
$\sigma_x
=\ket{\uparrow}\bra{\downarrow}+\ket{\downarrow}\bra{\uparrow}$. In the
state $\ket{\uparrow}$ ($\ket{\downarrow}$) the excess charge is
located on the upper (lower) dot. The energy difference $\epsilon$ and
the coupling $\Delta$ between these two states can be tuned by the gate
voltages $V_{g1}$ and $V_{g2}$ (see Fig.~\ref{fig:1} (a)). $\hat Q=e\hat
N=e\sum_id_i^{\dagger}d_i$ is the excess charge on the
quantum dot (QD) of the mesoscopic capacitor. The latter is described by the Hamiltonian
\begin{equation}\label{eq:3}
  H_D = \sum_i\varepsilon_i d_i^{\dagger}d_i +\frac{\hat Q^2}{2C_{\Sigma}}\,.
\end{equation}
Here the first term describes the unperturbed level spectrum while the
second term gives the Coulomb interaction. $C_{\Sigma}=(1/C_i+1/C_2+1/C_g)^{-1}$ is the total series
capacitance. Finally the QD of the capacitor is coupled to the lead via the
tunneling Hamiltonian $H_T=\sum_{ik}t_{ik}c_k^{\dagger}d_i+h.c.$,
where $t_{ik}$ is the tunneling matrix element between state $i$ of
the dot and state $k$ of the lead and can be tuned with the gate
voltage $V_{qpc}$ (see Fig.~\ref{fig:1} (a)). The lead, where we
neglect the electron electron interaction, is described by $H_L = \sum_kE_kc_k^{\dagger}c_k$.
The entire system is described by the Hamiltonian
\begin{equation}\label{eq:6}
  H = H_{qb}+H_D+H_L+H_T\,.
\end{equation}
If not for the tunneling term $H_T$, which changes the charge
on the dot of the capacitor, a qubit prepared in one of the eigenstates of $H_{qb}$ for
a given charge $Q$ would remain in this state under the time evolution. Because of $H_T$ however, the charge on
the dot fluctuates leading to a modulation in time of the level
splitting of the qubit. If this modulation is slow enough though, the qubit will remain
in an instantaneous eigenstate of $H_{qb}$ at all times. To derive the
necessary conditions for this to be true, we follow~\cite{Johansson:06}, and apply a unitary transformation onto $H$,
which diagonalizes $H_{qb}$ in each subspace of fixed $N$.
\begin{equation}\label{eq:12}
H^{\prime}={U(\hat N)}^{\dagger}HU(\hat N)\,.
\end{equation}
With $\eta_0={\rm
arccot}\left[\frac{\epsilon}{\Delta}\right]$, the unitary operator to
second order in the coupling $\kappa$ is explicitly given by
$U(\hat
N)=\hat a_0U_0
+ \hat a_1U_1$, with
\begin{equation}\label{eq:13}
  \hat a_0= 1-\frac{\kappa^2\Delta^2}{{8\Omega_0}^4}\hat N^2\quad\text{and}\quad \hat a_1= \frac{\kappa\Delta}{2\Omega_0^2}\hat N\left(1-\frac{\kappa\epsilon}{\Omega_0^2}\right)\,,
\end{equation}
where $\Omega_0=\sqrt{\epsilon^2+\Delta^2}$ is the bare Rabi frequency
and
\begin{equation}\label{eq:21}
  U_0=\begin{pmatrix}\cos\frac{\eta_0}{2}&-\sin\frac{\eta_0}{2}\\\sin\frac{\eta_0}{2}&\cos\frac{\eta_0}{2}\end{pmatrix},
  U_1=\begin{pmatrix}\sin\frac{\eta_0}{2}&\cos\frac{\eta_0}{2}\\-\cos\frac{\eta_0}{2}&\sin\frac{\eta_0}{2}\end{pmatrix}
\end{equation}
Note that ${U_0}^{\dagger}U_0={U_1}^{\dagger}U_1=\openone$ while
${U_0}^{\dagger}U_1=-{U_1}^{\dagger}U_0=i\sigma_y$. Using the fact
that $[[H_T,\hat N],\hat N]=H_T$, we finally obtain
\begin{equation}\label{eq:24}
U^{\dagger}H_TU =
{H_T}+i\sigma_y\frac{\kappa\Delta}{2\Omega_0^2}[H_T,\hat N]+O(\kappa^3)\,.
\end{equation}
where we have neglected a small $O(\kappa^2)$ renormalization of the tunneling
amplitudes $t_{ik}$, which is insensitive to the state of the qubit. In the linear response regime, the time scale on which $\braket{\hat N(t)}$
fluctuates is set by the
inverse of the drive frequency $\omega$. Therefore the energy
available for making a real transition between the qubit eigenstates,
which is given by the second term on
the right-hand side of Eq.~(\ref{eq:24}), is proportional to
$\hbar\omega\kappa\Delta/(2\Omega_0^2)$. Demanding that this energy be
small compared to the level splitting $\Omega_0$ of the qubit leads us
to the following adiabatic condition on the drive frequency
\begin{equation}\label{eq:2}
\hbar\omega \ll \frac{2\Omega_0^3}{\kappa\Delta}\,.
\end{equation}
Let us briefly discuss this condition. We see that for
$\Delta= 0$, we can drive the system as fast as we wish provided
$\epsilon\not=0$. This simply reflects the fact that for $(\Delta=0)\ll\epsilon$
the two eigenstates of the qubit, which in fact are the charge states
in this limit, are decoupled from one another. We also see that
the weaker the coupling, the faster we may drive the system without
inducing transitions, which is
intuitively reasonable. For realistic values of the parameters;
$\Delta=\Omega_0=5\,\mu{\rm  eV}$, $\epsilon=0$ and $\kappa=50\,\eta{\rm
  eV}$, we find $2\Omega_0^3/(\kappa\Delta)\gtrsim 1.5\cdot 10^{12}\,{\rm Hz}$, so that even for drive frequencies in the GHz regime we are still safely in the
adiabatic regime.

In the adiabatic approximation and for weak coupling, i.e. $\kappa\ll\Omega_0$, the
dynamics of the system is thus appropriately described to second order
in $\kappa$ by the purely longitudinal effective Hamiltonian $H_{{\rm eff}}=H_+\ket{+}\bra{+}+H_-\ket{-}\bra{-}$, where
\begin{equation}\label{eq:22}
H_{\pm} = \pm\frac{\Omega_0}{2} +
\sum_i\varepsilon_i^{\pm}d_i^{\dagger}d_i+\frac{e^2}{2C_{{\rm eff}}^{\pm}}\hat N^2+H_L+{H_T}\,.
\end{equation}
Here $\ket{\pm}$ are the adiabatic eigenstates
of $H_{qb}$. The presence of the qubit appears thus as a
renormalization of the spectrum of the QD of the detector:
$\varepsilon_i^{\pm}=\varepsilon_i\pm\kappa\epsilon/(2\Omega_0)$,
and a
renormalization of the geometric capacitance~\cite{Duty:05,Sillanpaa:05} of the dot vis-\`a-vis the
gate $V_{g1}$: $1/C_{{\rm
    eff}}^{\pm}=1/C_{\Sigma}\pm\kappa^2\Delta^2/(2\Omega_0^3)$.

Formally, the effective Hamiltonian we have just derived is exactly
the same as the one of a mesoscopic capacitor with a single level
spectrum $\varepsilon_i^{\pm}$ and a geometric capacitance
$C_{{\rm eff}}^{\pm}$. Within the self-consistent Hartree
approximation~\cite{Buttiker:93b,Brouwer:05a}, the linear response of
a mesoscopic capacitor to an applied
AC voltage is known~\cite{Buttiker:93b,Brouwer:05a}. For short RC
times $\tau_{RC}^{\pm}\equiv R_qC_{\mu}^{\pm}$ such that $\omega \tau_{RC}^{\pm}\ll 1$, the mesoscopic capacitor is equivalent to an RC circuit
with the impedance $Z_0^{\pm}(\omega) = R_q + i/(\omega C_{\mu}^{\pm})$.
Here $R_q$ is the charge relaxation resistance, which at zero temperature and for a single channel capacitor
is universal and given by half a resistance quantum,
i.e. $R_q=h/(2e^2)$. The electrochemical capacitance $C_{\mu}^{\pm}$
however depends on $C_{{\rm eff}}^{\pm}$ and on the density of states
(DOS) of
the capacitor and is thus sensitive to the state of the
qubit. Explicitly one finds~\cite{Buttiker:93b}
\begin{equation}
\frac{1}{C_{\mu}^{\pm}}=\frac{1}{C_{\rm
    eff}^{\pm}}+\frac{1}{e^2\nu_{\pm}(E_F)}\,.
\end{equation}
Here $\nu_{\pm}(E_F)$ is the DOS at the
Fermi-energy of the QD with the shifted spectrum
$\{\varepsilon_i^{\pm}\}$. The electrochemical capacitance thus acts like the pointer of a
measurement device. At the degeneracy point $\epsilon=0$, the shift of the levels
vanishes, while the correction to the capacitance is maximal. If to the contrary $\epsilon\gg\Delta$
then the correction to the capacitance vanishes while the dot spectrum is
maximally shifted by the amount $\pm\kappa/2$.

Let us now discuss a way of probing the
electrochemical capacitance in the
high frequency regime. Using a dispersive read-out scheme similar to
\cite{Schoelkopf:98,Johansson:06}, we embed the effective capacitor
into an LC tank-circuit and via a standard homodyne detection scheme~\cite{Gardiner:00},
probe the phase shift of waves reflected from the tank-circuit (see
Fig.~\ref{fig:1} (b)). It is important to note that in contrast to~\cite{Schoelkopf:98}, we here do not want to measure the
resistance of our effective capacitor. Indeed, owing to the
universality of the charge relaxation resistance in the single channel
limit, this quantity is actually insensitive
to the state of the qubit. Instead, we propose to detect the phase
shift of a reflected signal, which is determined by the
non-dissipative part of the response of the mesoscopic capacitor.

To second order in $C_{\mu}^{\pm}/C$, the shifted resonance frequency of the tank-circuit is
given by
\begin{equation}\label{eq:7}
\omega_{osc}^{\pm}\approx\omega_0\left(1-\frac{1}{2}\frac{C_{\mu}^{\pm}}{C}\right)-i\omega_0^2\frac{C_{\mu}^{\pm}}{2C}\tau_{RC}^{\pm}\,,
\end{equation}
where $\omega_0=1/\sqrt{LC}$ is the bare oscillator resonance frequency. Notice that because of the finite resistance $R_q$, the
oscillation of the LC circuit is damped. This is reflected in the
non-vanishing imaginary part of $\omega_{osc}^{\pm}$ in
Eq.~(\ref{eq:7}). In oder words, photons coming down the transmission
line toward the LC-tank circuit, will be dissipated with some finite
probability. The reflected photons however will experience a
phase shift, which depends on the state of the qubit. It is this phase
shift which we propose to measure.

The impedance of the tank-circuit which terminates the transmission
line is $Z_{\pm}(\omega)=iL({\omega_{osc}^{\pm}}^2-\omega^2)/\omega$.
From this, we can calculate the complex reflection coefficient
$\mathcal{R}^{\pm}$ of the
transmission line with characteristic impedance $Z_0$, relating
incoming and outgoing modes via
$a_{out}^{\pm}(\omega)=\mathcal{R}^{\pm}(\omega)a_{in}(\omega)$. We find
\begin{equation}\label{eq:9}
\mathcal{R}^{\pm}(\omega)=\frac{Z_0-Z_{\pm}(\omega)}{Z_0+Z_{\pm}(\omega)}=\frac{\omega^2-(\omega_{osc}^{\pm})^2-i\eta_0\omega}{\omega^2-(\omega_{osc}^{\pm})^2+i\eta_0\omega}\,,
\end{equation}
with $\eta_0 = Z_0/L$. Because $(\omega_{osc}^{\pm})^2$
has a non-vanishing imaginary part, $\mathcal{R}^{\pm}$ is not unitary. At
the bare resonance frequency, we obtain,
$\mathcal{R}_{\pm}(\omega_0) = \gamma_{\pm} e^{i\phi_{\pm}}$, with
\begin{equation}\label{eq:15}
\gamma_{\pm} =
1-2\frac{R_q}{Z_0}\left(\frac{C_{\mu}^{\pm}}{C}\right)^2+O\left(\left(C_{\mu}^{\pm}/C\right)^3\right)\,,
\end{equation}
and
\begin{equation}\label{eq:16}
\phi_{\pm} = Q_0\frac{C_{\mu}^{\pm}}{C}\left(2-\frac{1}{2}\left(\frac{C_{\mu}^{\pm}}{C}\right)\right)+O\left(\left(C_{\mu}^{\pm}/C\right)^3\right)\,.
\end{equation}
Here we have introduced the quality factor
$Q_0=\sqrt{L/C}/Z_0$ of the resonator plus
transmission line circuit. To leading order, the probability of a
photon to be dissipated is thus given by
$1-\gamma_{\pm}^2=4(R_q/Z_0)(C_{\mu}^{\pm}/C)^2$. Also, we remark that the
leading order correction to the reflection phase due to a finite $R_q$
is of order $(C_{\mu}^{\pm}/C)^3$. Finally note that the leading order correction to $\gamma_{\pm}$
is independent of $L$. This is ultimately the reason why we can
achieve a large Heisenberg efficiency; increasing $L$ increases the signal
without increasing the dissipation.

We next derive expressions
for the measurement and dephasing rates~\cite{Gardiner:00,Johansson:06,Clerk:09}.
The measured quantity is the number of photons reflected from the load in time $T$. By mixing this signal with a
strong signal from a local oscillator driven at the same frequency
$\omega_0$ as the drive and afterwards taking the average, the measured number of photons $n_{\pm}(T)$ becomes
sensitive to the reflection phase shift, which in turn depends on the
state of the qubit. The two eigenstates are said to be resolved, when the difference $\Delta n(T)=n_+(T)-n_-(T)$
becomes larger than the noise. The time when this happens defines the
measurement time $T_m$. Let us consider a monochromatic coherent state input
with amplitude $\beta_0$.
\begin{equation}
\ket{\psi}_{in}=\exp\left[T
    (\beta_0a^{\dagger}_L(\omega_0)-\beta_0^*a_L(\omega_0))\right]\ket{0}\,,
\end{equation}
where $a_L^{\dagger}(\omega)$ creates an incoming photon at
frequency $\omega$. Using the same definition for the signal to noise
ratio as in~\cite{Clerk:09},
we find a measurement rate given by
\begin{equation}\label{eq:29}
\Gamma_{m}\equiv {T_m}^{-1}=|\beta_0|^2\frac{(\gamma_++\gamma_-)^2}{\gamma_+^2+\gamma_-^2}\sin^2(\Delta\phi/2)\,,
\end{equation}
where $\Delta\phi=\phi_+-\phi_-$. We note that $\Gamma_m$ is bounded from above
by $2|\beta_0|^2$, or twice the photon injection
rate. Incidentally, this is the maximally achievable measurement rate in the absence
of dissipation, where $\gamma_{\pm}=1$.

To derive the dephasing rate, we essentially follow the quantum
information theoretic argument of~\cite{Clerk:09}
and adapt it to a dissipative system. The resistor is replaced
by a semi-infinite transmission line with characteristic impedance
$R_q$ (see Fig.~\ref{fig:1} (b)), which is then quantized~\cite{Yurke:84}. Hence we determine the transmission
coefficient $\mathcal{T}^{\pm}$ for photons to be dissipated. The measurement can be represented as the entangling process
\begin{equation}\label{eq:17}
(\alpha\ket{+}+\beta\ket{-})\ket{\beta_0}\rightarrow\alpha'\ket{+}\ket{\beta_+}+\beta'\ket{-}\ket{\beta_-}\,,
\end{equation}
where detector states after the scattering are given by a product of
phase shifted and damped coherent states as
\begin{equation}\label{eq:18}
\ket{\beta_{\pm}} =\ket{\beta_0 \gamma_{\pm}e^{i\phi_{\pm}}}\otimes \ket{\beta_0\sqrt{1-\gamma_{\pm}^2}e^{i\theta_{\pm}}}\,.
\end{equation}
Here $\theta_{\pm}=\arg(\mathcal T^{\pm})$
 is the phase shift of the dissipated
photons. The off-diagonal elements of the reduced density
matrix of the qubit are proportional to the overlap of the detector
states, i.e. $|\rho_{12}|\sim|\braket{\beta_+|\beta_-}|$. For long
times, these elements decay exponentially defining the dephasing rate
by $|\rho_{12}|\sim\exp[-\Gamma_{\phi}T]$. We find explicitly
\begin{equation}\label{eq:20}
\Gamma_{\phi}=|\beta_0|^2\left[1-D_1\cos(\Delta\phi)-D_2\cos(\Delta\theta)\right]\,,
\end{equation}
with $D_1=\gamma_+\gamma_-$ and $D_2=\sqrt{(1-\gamma_+^2)(1-\gamma_-^2)}$.
From Eqs.~(\ref{eq:29}) and (\ref{eq:20}) we finally obtain the Heisenberg
efficiency of our detector
\begin{equation}\label{eq:23}
\eta \equiv \frac{\Gamma_m}{\Gamma_{\phi}}=\frac{L/C}{L/C+R_qZ_0}+O\left((C_{\mu}^{\pm}/C)^2\right)\,.
\end{equation}
\begin{figure}[b]
\begin{center}
\includegraphics[width=.45\textwidth]{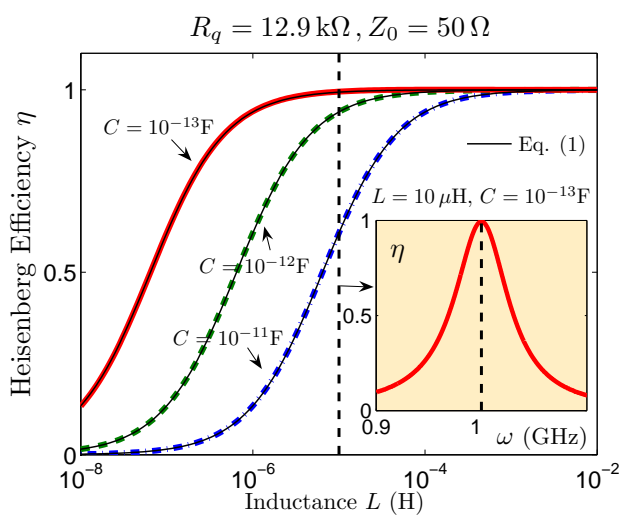}
\caption{(Color online) Efficiency $\eta$ as a function of inductance
  $L$. The
  inset shows $\eta(\omega)$ for $L=10\,{\mu\rm
    H}$ and $C=10^{-13}\,{\rm }F$.\label{fig:2}}
\end{center}
\end{figure}
Thus we find that to leading order the Heisenberg efficiency does not depend on
$C_{\mu}^{\pm}$. This result holds as long as $\omega_0\tau_{RC}^{\pm}\ll
1\ll C/C_{\mu}^{\pm}$. To reach acceptable
efficiency, we need to have $L/C\gg R_qZ_0$. Fig.~\ref{fig:2}
shows the Heisenberg efficiency as a function of $L$ for realistic parameters.
Decreasing $C$, which increases $\Delta\phi$, increases the efficiency and
at the same time increases the measurement frequency
$\omega_0=1/\sqrt{LC}$. For example for $L=10\,\mu \rm H$,
and $C= 100\, {\rm fF}$, we have $\omega_0=1\,{\rm GHz}$ and $\eta=99.4\%$
(see full thick (red) curves on Fig.~\ref{fig:2}).

In conclusion, we have shown that the mesoscopic capacitor can in
principle be operated as an efficient detector in the GHz regime. We find that to leading order its efficiency is universal, i.e. independent of the
microscopic details of the detector and qubit. This universality can
be directly traced back to the experimentally demonstrated~\cite{Gabelli:06} universality~\cite{Buttiker:93b} of the charge relaxation
resistance of a mesoscopic capacitor.

This work is supported by the Swiss NSF, MaNEP and
the STREP project SUBTLE.
\bibliography{Bibliography}
\bibliographystyle{apsrev}
\end{document}